\documentclass[a4paper]{jpconf}
\usepackage{ae,aecompl,aeguill} 
\usepackage{graphicx,color,setspace,amsmath}
\usepackage{amssymb}
\usepackage{mathrsfs}
\usepackage{lscape}
\usepackage{verbatim}
\usepackage{amsfonts}  
\usepackage{amsbsy}    

\begin{document}
\title{In search for the pairing glue in cuprates by non-equilibrium optical spectroscopy}

\author{F Cilento$^1$, S Dal Conte$^{2,3}$, G Coslovich$^4$, F Banfi$^{2,3}$, G Ferrini$^{2,3}$, H Eisaki$^5$, M Greven$^6$, A Damascelli$^{7,8}$, D van der Marel$^9$, F Parmigiani$^{1,10}$ and C Giannetti$^{2,3}$}

\address{$^1$Sincrotrone Trieste S.C.p.A., Basovizza I-34012, Italy.}
\address{$^2$I-LAMP (Interdisciplinary Laboratories for Advanced Materials Physics), Universit$\mathrm{\grave{a}}$ Cattolica del Sacro Cuore, Brescia I-25121, Italy.}
\address{$^3$Department of Mathematics and Physics, Universit$\mathrm{\grave{a}}$ Cattolica del Sacro Cuore, Brescia I-25121, Italy.}
\address{$^4$Materials Sciences Division, Lawrence Berkeley National Laboratory, Berkeley, CA 94720, USA.}
\address{$^5$Nanoelectronics Research Institute, National Institute of Advanced Industrial Science and Technology, Tsukuba, Ibaraki 305-8568, Japan}
\address{$^6$School of Physics and Astronomy, University of Minnesota, Minneapolis, MN 55455, USA.}
\address{$^7$Department of Physics and Astronomy, University of British Columbia, Vancouver, BC V6T 1Z1, Canada.}
\address{$^8$Quantum Matter Institute, University of British Columbia, Vancouver, BC V6T 1Z4, Canada.}
\address{$^9$D\'epartement de Physique de la Mati$\mathrm{\grave{e}}$re Condens\'ee, Universit\'e de Gen$\mathrm{\grave{e}}$ve, Gen$\mathrm{\grave{e}}$ve CH1211, Switzerland.}
\address{$^{10}$Department of Physics, Universit$\mathrm{\grave{a}}$ degli Studi di Trieste, Trieste I-34127, Italy.}

\ead{claudio.giannetti@unicatt.it}

\begin{abstract}
In strongly correlated materials the electronic and optical properties are significantly
affected by the coupling of fermionic quasiparticles to different degrees of freedom, such as lattice vibrations and bosonic excitations of electronic origin. Broadband ultrafast spectroscopy \cite{Cilento2010,Giannetti2011} is
emerging as the premier technique to unravel the subtle interplay between quasiparticles and electronic or phononic collective excitations, by their different characteristic timescales and spectral responses. 
By investigating the femtosecond dynamics of the optical properties of Bi$_{2}$Sr$_{2}$Ca$_{0.92}$Y$_{0.08}$Cu$_2$O$_{8+\delta}$ (Y-Bi2212) crystals over the 0.5-2 eV energy range, we disentangle the electronic and phononic contributions to the generalized electron-boson Eliashberg function \cite{vanHeumen2009,Carbotte2011}, showing that the spectral distribution of the electronic excitations, such as spin fluctuations and current loops,
and the strength of their interaction with quasiparticles can account for the
high critical temperature of the superconducting phase transition \cite{DalConte2012}. Finally, we discuss how the use of this technique can be extended to the underdoped region of the phase diagram of cuprates, in which a pseudogap in the quasiparticle density of states opens. 

The microscopic modeling of the interaction of ultrashort light pulses with unconventional superconductors will be one of the key challenges of the next-years materials science, eventually leading to the full understanding of the role of the electronic correlations in controlling the dynamics on the femtosecond timescale.
\end{abstract}

\section{Introduction}
\subsection*{The generalized bosonic function}
In conventional metals the scattering between quasiparticles (fermions) and lattice vibrations (bosons) is the microscopic mechanism that determines the transport and the optical properties. At low temperatures the electron-phonon interaction, denoted by $\alpha^2F(\Omega)$, provides the "glue" for the formation of the Cooper pairs, leading  to the instability of the Fermi-liquid ground state upon the formation of the superconducting condensate. The critical temperature of the superconducting phase transition, that is of the order of a few degrees, is determined by the electron-phonon coupling constant, i.e. $\lambda_{e-ph}$=$2\int\alpha^2F(\Omega)/\Omega\;d\Omega$, through the McMillan's formula \cite{Allen1975}.  

In the cuprate superconductors, the electronic correlations are responsible for the emergence of complex orders, influencing the phase diagram even far from the antiferromagnetic insulating phase at zero doping. These novel degrees of freedom, like paramagnon excitations \cite{LeTacon2011}, charge fluctuations \cite{Ghiringhelli2012} or loop currents \cite{Li2010} provide additional scattering channels to quasiparticles (QPs), strongly affecting their lifetime and dispersion over an energy range of the same order of the spectral distribution of the excitations. An "effective" description of the generic interaction between quasiparticles and excitations of both electronic and phononic nature, whose distribution at a given temperature follows the Bose-Einstein statistics, can be obtained by replacing the $\alpha^2F(\Omega)$ with a more general electron-boson coupling function \cite{Carbotte2011}, i.e., the \textit{Bosonic Function} $\Pi(\Omega)$, defined as:
\begin{equation}\label{TotalGlue}
	\Pi(\Omega) \equiv \alpha^2F(\Omega)+I^2 \chi(\Omega)
\end{equation}
where $I^2 \chi(\Omega)$ accounts for the coupling with all the bosonic excitations of electronic origin. 

The signatures of the QP-boson interaction usually manifest in most of the experiments that probe the electronic properties at equilibrium. The generalized bosonic function $\Pi(\Omega)$ can be extracted \cite{Carbotte2011} by analyzing the kinks in angle-resolved photoemission data \cite{Damascelli2003}, the dip features in tunneling spectra 	\cite{Lee2006,Ahmadi2011}, the frequency-dependent scattering rate in optical spectroscopy \cite{Hwang2007,vanHeumen2009} and Raman spectra \cite{Muschler2010}. $\Pi(\Omega)$ shows some ubiquitous features, independent of the technique used, like a strong peak at 50-80 meV and a broad continuum that extends up to 300-400 meV. Although a main effort has been recently made \cite{Carbotte2011}, the determination of the relative weight of the electronic and phononic contributions to $\Pi(\Omega)$ remains elusive, since the electronic and phononic excitations coexist on the 0-90 meV energy scale. Solving this problem would constitute a step forward to unravel the puzzle of superconductivity in cuprates, while addressing the major question whether or not high-temperature superconductivity can be described in terms of a generalized Eliashberg formalism, in which the attractive interaction is 'retarded', i.e., mediated by virtual bosonic excitations of novel origin \cite{Millis1990}.

\begin{figure}[t]
\begin{center}
\includegraphics[bb= 60 30 620 400, width=25pc]{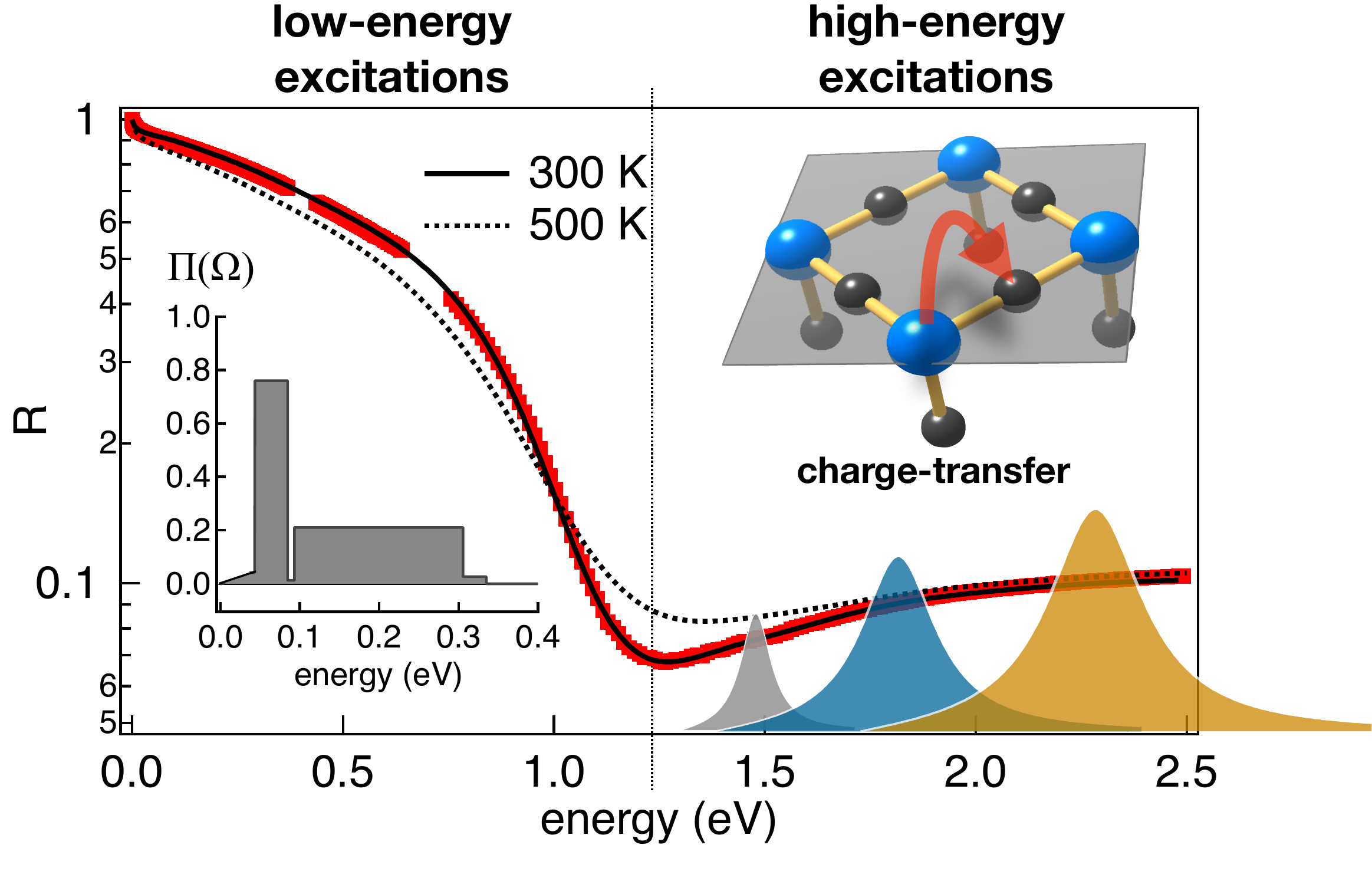}
\end{center}
\caption{\label{figure1}Reflectivity of the Bi$_{2}$Sr$_{2}$Ca$_{0.92}$Y$_{0.08}$Cu$_2$O$_{8+\delta}$ crystal at optimal doping (T$_c$=96 K), as measured by conventional spectroscopic ellipsometry (red dots). The reflectivity is well reproduced by an extended Drude model and a sum of Lorentz oscillators (black line). The total bosonic glue, $\Pi(\Omega)$, extracted from the data is shown in the left inset. The dashed black line represents the reflectivity calculated for the same system at T=500 K.}
\end{figure}

Non-equilibrium optical spectroscopy is emerging as a new tool to tackle these fundamental questions \cite{Cilento2010,Giannetti2011}. The founding concepts of this technique rely on the use of an ultrashort light pulse (pump) to prepare the system in a non-equilibrium state, i.e., with the distribution of fermionic QPs decoupled from the distribution of bosonic excitations. The induced transient changes in the optical properties can be modeled by a two-steps process: i) the sudden photo-injection of fermionic QPs results in an effective increase of the plasma frequency of the Drude peak in the optical conductivity, without any change of the QPs scattering rate; ii) the subsequent heating of the bosonic excitations induces an increase of the scattering rate and a broadening of the Drude peak, without any change of the plasma frequency. 

Since changes in the plasma frequency and in the scattering rate affect the optical conductivity in qualitatively different ways, optical spectroscopy with femtosecond time-resolution can be used to probe the temporal evolution of the distribution of fermionic QPs and bosonic excitations, as a function of the delay from the pump pulse. Furthermore, the dynamics of bosonic excitations of electronic and phononic nature can be disentangled on the basis of their different timescales while exchanging energy with QPs.

In this work, we will briefly introduce the basics of equilibrium and non-equilibrium optical spectroscopy, we will review the main results on the determination of the relative electronic and phononic contributions to $\Pi(\Omega)$ and, finally, we will show how this technique can be extended to investigate the opening of a pseudogap in the quasiparticle density of states and possible anomalies in the temperature dependance of the QP self-energy.

\section{Optical spectroscopy at equilibrium}
Optical spectroscopy at equilibrium is a fundamental tool to investigate the electronic properties of strongly correlated materials \cite{Basov2010} and unconventional superconductors \cite{Basov2005}. Since the dielectric function contains direct information about the scattering rate of QPs and their dynamical effective mass, optical spectroscopies are intrinsically sensitive to the coupling of QPs to bosonic excitations.
\subsection{The optical properties of a prototypical cuprate}
In Figure 1 we report  the  \textit{ab}-plane reflectivity of optimally-doped Bi$_{2}$Sr$_{2}$Ca$_{0.92}$Y$_{0.08}$Cu$_2$O$_{8+\delta}$ (Y-Bi2212) high-quality crystals \cite{Eisaki2004} ($T_c$=96 K), measured by conventional spectroscopic ellipsometry \cite{vanderMarel2003} at 300 K.
The normal-incidence reflectivity $R(\omega,T)$ is related to the dielectric function $\epsilon(\omega,T)$ by the relation: 
\begin{equation}
R(\omega,T)=\left| \frac{1-\sqrt{\epsilon(\omega,T)}}{1+\sqrt{\epsilon(\omega,T)}}\right|^2
\end{equation}
The dielectric function is obtained by applying the Kramers-Kronig relations to the reflectivity for 6 meV$<$$\hbar\omega$$<$0.74 eV and directly from ellipsometry for 0.2 eV$<$$\hbar\omega$$<$4.5 eV. This combination allows a very accurate determination of $\epsilon(\omega)$ in the entire combined frequency range. 

The reflectivity of Y-Bi2212 shows some general features common to most of cuprates:\\
i) below 1.25 eV the optical properties are dominated by a broad peak related to the optical response of the low-energy excitations in the conduction band. Interestingly, the dielectric function in this energy range cannot be simply reproduced by a Drude peak, in which a constant scattering rate $\tau$ and effective mass $m^*$ of the QPs is assumed. The strongly frequency-dependent scattering rate and effective mass, resulting from the interaction with bosonic excitations, will be accounted for by the more general extended Drude model \cite{Basov2005}, presented in the next section;\\ 
ii) above 1.25 eV, the high-energy interband transitions dominate. In this energy range, the equilibrium dielectric function can be modeled as a sum of Lorentz oscillators at $\sim$1.5, 2, 2.7 and 3.9 eV \cite{Giannetti2011}. The attribution of these interband transitions in cuprates is a subject of intense debate. The ubiquitous charge-transfer (CT) gap edge (hole from the upper Hubbard band with $d_{x^2-y^2}$ symmetry to the O-2$p_{x,y}$ orbitals) in the undoped compounds is about 2 eV \cite{Basov2005}. Upon doping, a structure reminiscent of the CT gap moves to higher energies, while the gap is filled with new transitions. This trend has been recently reproduced by Dynamical Mean Field Theory (DMFT) calculations of the electron spectral function and of the \textit{ab}-plane optical conductivity for the hole-doped three-band Hubbard model \cite{DeMedici2009}. The structures appearing in the dielectric function at 1-2 eV, that is, below the remnant of the CT gap at 2.5-3 eV, are possibly related to transitions between many-body Cu-O states at binding energies as high as 2 eV (for example singlet states) and states at the Fermi energy;\\
iii) the dressed plasma frequency, $\bar{\omega_p}$, defined through the relation Re$\{\epsilon(\bar{\omega_p},T)\}$=0, is $\sim$1 eV. In this energy range, the signatures of the pump-induced change of the QP distribution and of the heating of bosons can be more effectively disentangled, as will be discussed in Section 3. 

\subsection{The extended Drude model}
In the extended Drude model (EDM) the physical processes responsible for the renormalization of the lifetime and effective mass  of the QPs are accounted for in a phenomenological way, by replacing the frequency-independent scattering time $\tau$ with a complex temperature- and frequency-dependent scattering time $\tau(\omega,T)$:
\begin{equation}
\tau^{-1} \Rightarrow \tilde{\tau}^{-1}(\omega)=\tau^{-1}(\omega)-i\omega\tilde{\lambda}(\omega)=-iM(\omega,T)
\end{equation}
where $1+\tilde{\lambda}(\omega)=\frac{m^*}{m}(\omega)$ is the mass renormalization of the QPs due to many-body interactions and $M(\omega,T)$ is the memory function.
\\
In the EDM, the dielectric function $\epsilon_D(\omega,T)$ is given by:
\begin{equation}\label{EDM_Drude1}
\begin{split}
\epsilon_D(\omega,T)=1-\frac { {\omega_{p}}^{2}} {\omega(\omega+ M(\omega,T))}
=1-\frac { {\omega_{p}}^{2} } {\omega(\omega(1+\tilde{\lambda}(\omega,T))+ i/\tau(\omega,T))}
\end{split}
\end{equation}
while the Drude optical conductivity $\sigma_D(\omega,T)$=$[1-\epsilon_D(\omega,T)]i\omega/4\pi$ reads:
\begin{equation}\label{EDM_Drude2}         
\begin{split}                                                                                                                                                                                                                                                                                                                                                                                                                                                                                                                                                                                                                                                                                                                                                                                                                                                                                                                                                                                                                                                                                                                                                                                                                                                                                                                                                                                                                                                                                                                                                                                                                                                                                                                                                                                                                                                                                                                                                                                                                                                                                                                                                                                                                                                       
\sigma_D(\omega,T)=\frac{i}{4\pi} \frac { {\omega_{p}}^{2} } {\omega+M(\omega,T)}
=\frac{1}{4\pi} \frac {{\omega_{p}}^{2}} {1/\tau(\omega,T)-i\omega(1+\tilde{\lambda}(\omega,T))}
\end{split}
\end{equation}
\begin{figure}[t]
\begin{minipage}{22pc}
\includegraphics[width=22pc]{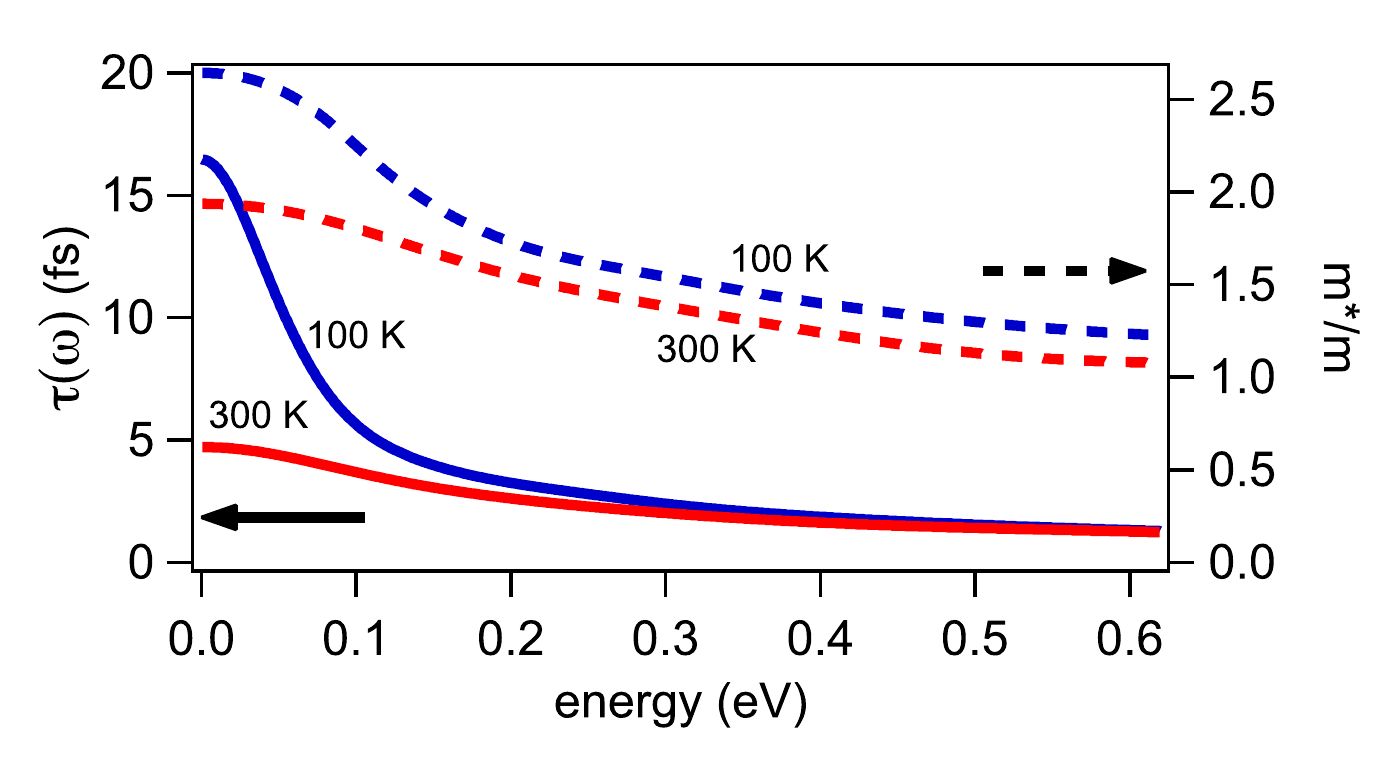}
\end{minipage}\hspace{0.5pc}
\begin{minipage}{15pc}
\caption{\label{lifetime}Frequency-dependent scattering rate (left axis) at 300 K (red line) and 100 K (blue line) and effective mass (right axis) at 300 K (red dashed line) and 100 K (blue dashed line) of the fermionic quasiparticles.}
\end{minipage}
\end{figure}
\\
The renormalized scattering rate and effective mass can be directly extracted from the measured Drude optical conductivity, through the relations:
\begin{eqnarray}
1/\tau(\omega,T)&=&\frac{\omega_p^{2}}{4\pi}\mathrm{Re} \left ( \frac{1}{\sigma_D(\omega,T)} \right )\\
1+\tilde{\lambda}(\omega,T)&=&-\frac{\omega_p^{2}}{4\pi} \frac{1}{\omega}\mathrm{Im} \left ( \frac{1}{\sigma_D(\omega,T)} \right)
\end{eqnarray}
\\
Although this phenomenological version of the EDM does not provide any clue about the microscopic mechanisms responsible for the renormalization of the energy dispersion and lifetime of the QPs, it is very useful for directly extracting $\tau(\omega,T)$ and $m^*(\omega,T)/m$ from the optical data. In Figure 2, we report $\tau(\omega,T)$ (solid lines, left axis) and $m^*(\omega,T)/m$ (dashed lines, right axis) obtained by applying Eqs. 6-7 to the low-energy part of $\sigma_D(\omega,T)$, measured on optimally-doped Y-Bi2212 samples. At T=300 K (red lines), the value of the QP scattering time is $\sim$5 fs at very low frequencies and decreases towards an asymptotic value of $\sim$2 fs above 0.5 eV. These values of  $\tau(\omega,T)$ imply that QPs scatters very quickly and exchange energy with bosons on the very femtosecond timescale. As the temperature decreases, the opening of the pseudogap prevents QPs from being scattered and an increase of the scattering time below 100 meV is measured, while the value of $\tau(\omega,T)$ above 100 meV is almost unaltered. A similar behavior is measured for the effective mass, that decreases towards the $m^*/m$=1 asymptotic value above 0.5 eV.

\subsection{The extraction of the bosonic function from the optical conductivity}
From the microscopic point of view, the Extended Drude formalism can be derived from the Holstein theory for normal metals \cite{Kaufmann1998}. Considering the Kubo formula and using complex diagrammatic techniques to evaluate the electron and boson thermal Green's functions and omitting vertex corrections (Migdal approximation), the Memory function $M(\omega,T)$ results:
\begin{equation}\label{Memory_EDM2_b}
M(\omega,T)=\omega\left\{ \int_{-\infty}^{+\infty} \frac{f(\xi,T)- f(\xi+\omega,T)}{\omega+\Sigma^*(\xi,T)-\Sigma(\xi+\omega,T)+i\gamma_{\mathrm{imp}}} d\xi \right\}^{-1}-\omega
\end{equation}
where $f$ is the Fermi-Dirac distribution, $\Sigma(\omega,T)$ and $\Sigma^*(\omega,T)$ the electron and hole \textbf{k}-space averaged self-energies and $\gamma_{\mathrm{imp}}$ an intrinsic decay rate that accounts for the scattering by impurities. We pinpoint that, although the memory function $M(\omega,T)$ has the same analytical properties of the single-particle self-energy $\Sigma(\omega,T)$, it has a conceptually different meaning, since the optical transition at frequency $\omega$ involves a particle-hole excitation of the many-body system and provides information about the joint particle-hole density of states.

The electron self-energy $\Sigma(\omega,T)$ can be calculated as a convolution integral between the bosonic function $\Pi(\Omega)$ and a kernel function $L(\omega,\Omega,T)$:
\begin{equation}\label{ESelf_EDM2}
\Sigma(\omega,T)=\int_0^\infty\Pi(\Omega)L(\omega,\Omega,T)d\Omega
\end{equation}

The kernel function
\begin{equation}\label{Kernel_EDM2_a}
L(\omega,\Omega,T)=\int \left[ \frac{n(\Omega',T)+f(\Omega,T)}{\Omega-\omega+\Omega'+i\delta} + \frac{1+n(\Omega',T)-f(\Omega,T)}{\Omega-\omega-\Omega'-i\delta} \right ]d\Omega'
\end{equation}
accounts for the distribution of the bosonic excitations through the Bose-Einstein distribution $n(\Omega,T)$, and can be calculated analytically:
\begin{equation}\label{Kernel_EDM2_b}
L(\omega,\Omega;T_e,T_b)=-2\pi i\left[n(\Omega,T_b)+\frac{1}{2}\right] +\Psi \left(\frac{1}{2}+i\frac{\Omega-\omega}{2\pi T_e}\right) -\Psi \left(\frac{1}{2}-i\frac{\Omega+\omega}{2\pi T_e}\right)
\end{equation}
where $\Psi$ are digamma functions and the dependence of the different terms on the temperatures of the electronic QPs ($T_e$) and bosonic excitations ($T_b$) has been made explicit. In this formalism, the frequency-dependent scattering rate is a consequence of the microscopic interaction of QPs with a distribution of bosons at temperature $T_b$.

Close to optimal doping, the vertex corrections beyond Eliashberg theory can be reliably neglected and the EDM can be safely used to extract $\Pi(\Omega)$ from the optical conductivity, measured at the equilibrium temperature $T$. Although sophisticated maximum entropy techniques \cite{Hwang2007} have been developed to unveil the rich details of the bosonic function, the main features can be evidenced by simply fitting the model in (5) to the experimental reflectivity $R(\omega,T)$, reported in Figure 1, and assuming a simple histogram form for $\Pi(\Omega)$. The resulting bosonic function is reported in the left inset of Figure 1 and is characterized by:\\
i) a low-energy part (up to 40 meV) linearly increasing with the frequency. This part is compatible with either the coupling of QPs to acoustic \cite{Johnston2011} and Raman-active optical \cite{Kovaleva2004} phonons or the linear susceptibility expected for a Fermi liquid \cite{Mirzaei2012};\\
ii) a narrow, intense peak centered at $\sim$60 meV, attributed to the anisotropic coupling to either out-of-plane buckling and in-plane breathing Cu-O optical modes \cite{Devereaux2004} or bosonic excitations of electronic origin such as spin fluctuations \cite{Dahm2009};\\
iii) a broad continuum extending up to 350 meV, well above the characteristic phonon cutoff frequency ($\sim$90 meV) and usually attributed to the coupling with spin fluctuations \cite{Abanov2001,Chubukov2005,Norman2006,LeTacon2011} or loop currents \cite{Varma2010}.

Using a sum of the EDM and four Lorentz oscillators at $\sim$1.5, 2, 2.7 and 3.9 eV, the reflectivity of optimally-doped Y-Bi2212 can be satisfactorily reproduced up to 2.5 eV photon energy (see the black solid line in Figure 1). In the EDM the main role of the temperature is to change the density of the bosons and, as a consequence, the scattering rate of QPs. In Figure 1 we report $R(\omega,T)$ calculated for $T$=500 K (dashed line). The $T$-related increase of the scattering rate induces a broadening of the Drude peak, resulting in a decrease of the reflectivity below  $\bar{\omega_p}$=1 eV and an increase of the reflectivity above $\bar{\omega_p}$. 

Although very useful to determine the microscopic origin of the frequency-dependent electronic properties and capable of reproducing most of the features of the temperature-dependance of the dielectric function, the EDM does not allow to evaluate the relative contributions of the phononic and electronic contributions to the total $\Pi(\omega)$, since they spectrally coexist below 90 meV.                                                                                                                                                                                                                                                                                                                                                                                                                                                                                     

\section{Non-equilibrium optical spectroscopy}
Non-equilibrium optical spectroscopy permits to monitor the change of the dielectric function and, in particular, of the QPs scattering rate, during the relaxation process after the interaction with an ultrashort optical light pulse. Adding the temporal ($t$) dimension to the frequency ($\omega$) dimension will be the key to disentangle the phononic and electronic contributions to $\Pi(\omega)$.
A sketch of the technique is reported in Figure 3. A 100 fs infrared optical pulse (1.5 eV photon energy) is used to excite the sample. The dynamics of the dielectric function is probed by combining two ultrafast optical techniques: i) supercontinuum light generation in a photonic fiber and simultaneous detection of the 1-2 eV spectral region through a linear array, after dispersion in a prism; ii) Optical Parametric Amplification providing tunable and short (<100 fs) output pulses in the 0.5-1.1 eV range. The details of the experiment can be found in Refs. \cite{Cilento2010,Giannetti2011,DalConte2012}.

\subsection{The extended Drude model out of equilibrium}
During the interaction with the pump pulse, two physical processes are expected to take place on the sub-picosecond timescale:\\
-High-energy fermionic excitations are impulsively injected by the pump pulse. Since the lifetime of these excitations is very short, they rapidly relax toward the Fermi energy, producing a high-density of QPs at lower excitation energies. As far as no pseudogap is present, which would provide constraints for the QPs scattering processes, we assume that the excess of QPs can be described through an effective increase $\delta T_e$ of the equilibrium electronic temperature $T_e$.\\
-The temperature $T_b$ of the distribution of the bosonic excitations shows a $\delta T_b$ increase because of the QP-boson coupling.\\ 
By noting that the Kernel function $L(\omega,\Omega;T_e,T_b)$ is the sum of separate terms depending on the electronic and bosonic temperatures, it is possible to use the EDM described in Section 2.3 to calculate the changes in the optical properties expected at $t$=0, i.e., soon after the excitation, in the two limits:
a) Non-thermal scenario, i.e., $\delta T_e$$\gg$$\delta T_b$, in which the QPs are almost completely decoupled from the bosons.
b) Quasi-thermal scenario, i.e., $\delta T_e$$\simeq$$\delta T_b$, in which the QPs and bosons are in thermal equilibrium at an effective temperature larger than the equilibrium temperature.

To facilitate the comparison with the experimental outcomes, we calculated the relative variation of the reflectivity at $t$=0, defined as the relative difference of the reflectivities out of equilibrium ($R_{oeq}$) and at equilibrium ($R_{eq}$):
\begin{equation}
\frac{\delta R}{R}(\omega,T)=\frac{R_{oeq}(\omega,T)-R_{eq}(\omega,T)}{R_{eq}(\omega,T)}=\frac{R(\omega,T_e+\delta T_e,T_b+\delta T_b)-R(\omega,T_e,T_b)}{R(\omega,T_e,T_b)}
\end{equation}
The results are reported in Figure 4. In the case $\delta T_e$$\gg$$\delta T_b$ (yellow solid line), the variation of the reflectivity shows a positive, intense feature just below $\bar{\omega_p}$=1 eV, while it is negligible at higher energies. This behavior can be rationalized in terms of a small increase of the plasma frequency without any change in the scattering rate $\tau(\omega)$. For the case $\delta T_e$$\simeq$$\delta T_b$ (black solid line), the reflectivity variation is dramatically different, since it is dominated by the increase of the density of the bosons and of the QPs scattering rate, as evidenced by the broadening of the Drude peak. In the quasi-thermal scenario, $\delta R/R(\omega, T)$ turns from positive to negative at $\bar{\omega_p}$.

These results open the way for probing the dynamics of the dielectric function over a wide energy range, above and below $\bar{\omega_p}$, allowing to discriminate between the non-thermal ($\delta T_e$$\gg$$\delta T_b$) and quasi-thermal ($\delta T_e$$\simeq$$\delta T_b$) scenarios and directly measure the time necessary for the heating of the bosonic excitations. 

\begin{figure}[t]
\begin{minipage}{20pc}
\includegraphics[bb= 140 340 420 480, width=18pc]{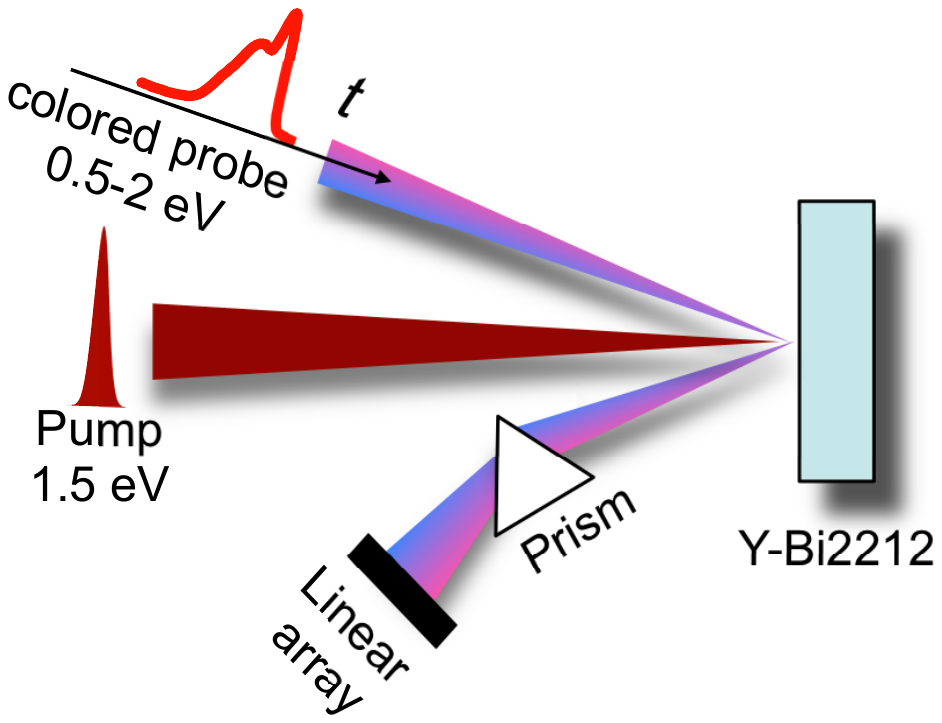}
\end{minipage}\hspace{1.5pc}
\begin{minipage}{15pc}
\caption{\label{setup}Sketch of the experimental setup for the 1.5 eV-pump and colored-probe optical measurements.}
\end{minipage} 
\end{figure}

\subsection{The bosonic function and the relaxation dynamics}
The key point to analyze the time-resolved data is that the same bosonic function $\Pi(\Omega)$ obtained from the extended Drude model (see Eq. 9) controls the temporal dynamics of the energy exchange between QPs and bosons, as pointed out by P. Allen in Ref. \cite{Allen1987}. This model, introduced for simple metals and referenced as the "effective-temperature model", can be extended to capture the more complex physics of cuprates, including the strong coupling with some Cu-O optical phonons. Formally, the  total bosonic function can be written as $\Pi(\Omega)$=$I^2 \chi(\Omega)$+$\alpha^2F(\Omega)_{SCP}$+$\alpha^2F(\Omega)_{lat}$, where $I^2 \chi(\Omega)$ refers to the bosonic excitations of electronic origin at the effective temperature $T_\mathrm{be}$, $\alpha^2F(\Omega)_{SCP}$ to the small fraction of strongly-coupled phonons (SCPs), e.g., buckling and breathing Cu-O optical modes,  at $T_\mathrm{SCP}$\cite{Perfetti2007} and $\alpha^2F(\Omega)_{lat}$  to all other lattice vibrations at $T_{lat}$, including acoustic and Raman-active optical phonons.

A set of four coupled differential equations can be used to represent the following physical processes: a short laser pulse, with power density (absorbed) $p$, impulsively increases the effective electronic temperature of the QPs with a specific heat $C_{e}$=$\gamma_e$$T_e$ ($\gamma_e$=$\pi^2N_cN(\epsilon_F)k^2_b/3$, $N_c$ being the number of cells in the sample and $N(E_F)$ the density of states of both spins per unit cell). $T_e$ will then relax through the energy exchange with all the coupled degrees of freedom  that linearly contribute to the total $\Pi(\Omega)$=$I^2 \chi(\Omega)$+$\alpha^2F(\Omega)_{SCP}$+$\alpha^2F(\Omega)_{lat}$.
The rate of the energy exchange among the different populations is given by \cite{Allen1987}:
\begin{eqnarray}
\frac{\partial T_e}{\partial t}&=&\frac{G(I^2 \chi,T_{be},T_{e})}{\gamma_eT_e}+\frac{G(\alpha^2F_{SCP},T_{SCP},T_{e})}{\gamma_eT_e}+\frac{G(\alpha^2F_{lat},T_{lat},T_{e})}{\gamma_eT_e}+\frac{p}{\gamma_e T_e}\\
\frac{\partial T_{be}}{\partial t}&=&-\frac{G(I^2 \chi,T_{be},T_{e})}{C_{be}}\\
\frac{\partial T_{SCP}}{\partial t}&=&-\frac{G(\alpha^2F_{SCP},T_{SCP},T_{e})}{C_{SCP}}\label{4TM3}\\
\frac{\partial T_{lat}}{\partial t}&=&-\frac{G(\alpha^2F_{lat},T_{lat},T_{e})}{C_{lat}}\label{4TM4}
\end{eqnarray}

where 
\begin{equation}
G(\Pi_{i},T_{i},T_{e})=\frac{6\gamma_e}{\pi \hbar k^2_b}\int^{\infty}_0d\Omega \Pi_{i}(\Omega) \Omega^2 [n(\Omega,T_{i})-n(\Omega,T_{e})]
\end{equation}
with $\Pi_{i}$=$I^2 \chi$, $\alpha^2F_{SCP}$, $\alpha^2F_{lat}$ and $n(\Omega,T_i)$=($e^{\Omega/k_BT_i}-1$)$^{-1}$ the Bose-Einstein distribution at the temperatures $T_{i}$=$T_{be}$, $T_{SCP}$, $T_{lat}$. The specific heat ($C_{SCP}$) of SCPs is proportional to their density of states and is taken as a fraction $f$ of the total specific heat, i.e., $C_{SCP}$=$fC_{lat}$.

Combining the four temperature model (4TM, Eqs. 13-17) with the extended Drude model (Eqs. 5,8,9,11), provides a full picture of the dynamics of $\delta R/R(\omega,T_{be},T_{SCP},T_{lat})$, in which $T_{be}(t)$, $T_{SCP}(t)$ and $T_{lat}(t)$  are functions of the time variable $t$ and their dynamics is controlled by the three different ratios $G(\Pi_\mathrm{i},T_\mathrm{i},T_\mathrm{e})/C_\mathrm{i}$, where $C_\mathrm{i}$=$C_\mathrm{be}$, $C_\mathrm{SCP}$ and $C_\mathrm{lat}$.

\section{Disentangling the electronic and phononic contributions to the bosonic function}

\begin{figure}[t]
\begin{center}
\includegraphics[bb= 220 40 560 520, width=15pc]{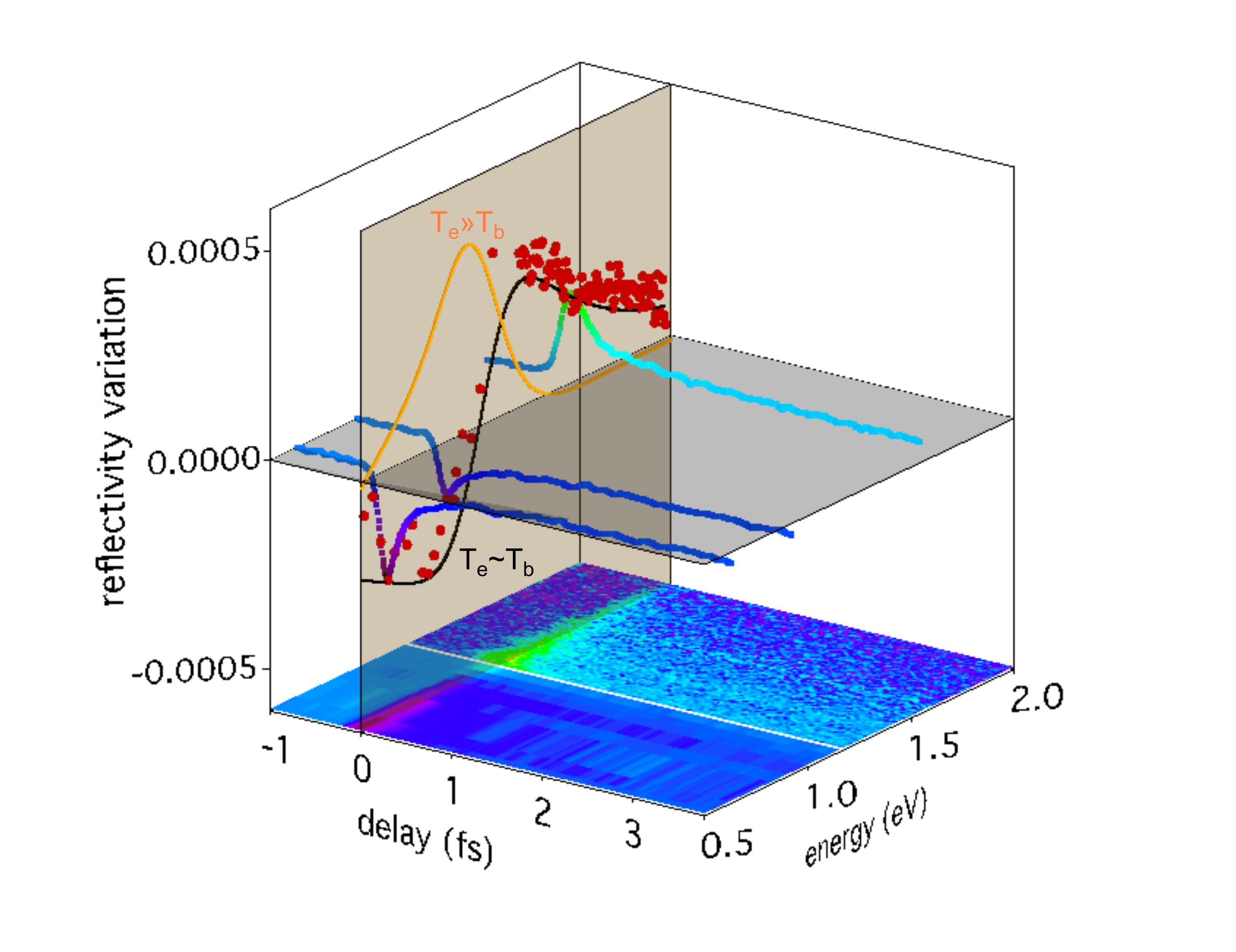}\hspace{2pc}%
\caption{\label{data}Relative reflectivity variation, i.e. $\delta R/R(\omega,t)$=($R_{exc}(\omega,t)$-$R_{eq}(\omega)$)/$R_{eq}(\omega)$, as a function of the probe photon energy and delay between the pump and probe pulses. The data have been taken on an optimally-doped Y-Bi2212 sample at room temperature. For further details, see Ref. \cite{DalConte2012}. The maximum $\delta R/R(\omega,t)$ at $t$=0 is calculated in the quasi-thermal (black line; $T_e$$\simeq$$T_b$) and non- thermal (yellow line; $T_e$$\gg$$T_b$) scenarios, using the parameters obtained from the fit to the equilibrium measurements. The red dots are the $\delta R/R(\omega,t$=$0)$ measured by time-resolved optical spectroscopy.}
\end{center}
\end{figure}

Figure 4 reports the $\delta R/R(\omega,t)$, as measured through non-equilibrium spectroscopy on optimally-doped Y-Bi2212 samples at T=300 K, versus time ($t$, $x$-axis) and frequency ($\omega$, $y$-axis). Some worthwhile features emerge from the data.\\
Already on the very short timescale (<100 fs), $\delta R/R(\omega,t)$ is negative below 1 eV and positive above 1 eV. The red dots are the values of $\delta R/R(\omega,t=0)$, i.e., the maximum reflectivity variation versus the probe wavelength. The data are well reproduced by the reflectivity variation calculated in quasi-thermal conditions ($\delta T_e$$\simeq$$\delta T_b$) through the EDM. It is worth noting that no signature of a change in the plasma frequency, characteristic of the non-thermal scenario ($\delta T_e$$\gg$$\delta T_b$), is detected within the time resolution of the experiment. Hence, it is possible to conclude that, already on the 100 fs timescale, the QPs are thermalized with some bosonic excitations participating to $\Pi(\Omega)$. The fast timescale of this thermalization implies a very large coupling and a relatively small specific heat. This finding strongly suggests that this process involves bosonic excitations of electronic origin and is consistent with the scattering time $\tau(\omega)$$\sim$1-5 fs estimated from the equilibrium optical conductivity measurements reported in Figure 2.\\
The temporal dynamics at fixed wavelength, i.e., $\delta R/R(t)$, is characterized by two decay times ($\sim$200 fs and $\sim$1 ps), universally measured in time-resolved experiments \cite{Kabanov1999,Gedik2005}. The faster dynamics is related to the coupling of QPs to SCPs, controlled by the $\alpha^2F_{SCP}$ part of the bosonic function and by $C_{SCP}$, while the slower dynamics is attributed to the coupling of QPs to all other lattice modes and it is controlled by $\alpha^2F_{lat}$ and $C_{lat}$.

A quantitative analysis of the data can be carried out by using the combination of the four temperature model (4TM, Eqs. 13-18) and of the extended Drude model (Eqs. 5,8,9,11), previously discussed. The different contributions to $\Pi(\Omega)$ are extracted by fitting the calculated $\delta R/R(\omega,t)$ to the time- and frequency-resolved reflectivity data, reported in Figure 4.  Considering that the energy distribution of phonons is limited to $<$90 meV, we assume that, for $\Omega$$>$90 meV, $\Pi(\Omega)$$\simeq$$I^2 \chi(\Omega)$. Within this assumption, the functional dependence of $\delta R/R(\omega,t)$ on $I^2 \chi(\Omega)$, $\alpha^2F(\Omega)_{SCP}$ and $\alpha^2F(\Omega)_{lat}$ is simplified as a parametric dependence on three coefficients $p_i$, where $I^2 \chi(\Omega)$=$p_{1}$$\Pi(\Omega$$<$90 meV)+$\Pi(\Omega$$>$90 meV), $\alpha^2F(\Omega)_{SCP}$=$p_{2}$$\Pi(\Omega$$<$90 meV) and $\alpha^2F(\Omega)_{lat}$=$p_{3}$$\Pi(\Omega$$<$90 meV). Considering the constraint that $\Pi(\Omega)$=$I^2 \chi(\Omega)$+$\alpha^2F(\Omega)_{SCP}$+$\alpha^2F(\Omega)_{lat}$ ($\Pi(\Omega)$ being the total glue function extracted from equilibrium optical spectroscopy and reported in Figure 1) and fixing the values $C_e$/$T_e$=$\gamma_e$=10$^{-4}$ Jcm$^{-3}$K$^{-2}$ and $C_{lat}$=2.27 Jcm$^{-3}$K$^{-1}$, 
the parameters phase-space of the model is significantly narrowed, allowing to unambiguously haul out the different contributions to $\Pi(\Omega)$ and to estimate $C_{be}$ and $C_{SCP}$. For further details, see Ref. \cite{DalConte2012}.

The analysis of the $\delta R/R(\omega,t)$ demonstrates that the entire high-energy part and $\sim$46$\%$ of the peak at $\sim$60 meV instantaneously thermalize with the QPs at a temperature $T_\mathrm{be}$$\simeq$$T_\mathrm{e}$. The spectral distribution and the estimated value of the specific heat of these excitations ($C_\mathrm{be}$$<$0.1$C_\mathrm{e}$) support their electronic origin. The $\alpha^2F(\Omega)_{SCP}$ is estimated to be $\sim$34$\%$ of the peak at $\sim$60 meV, corresponding to $\sim$17$\%$ of the total $\Pi(\Omega)$, while $\alpha^2F(\Omega)_{lat}$ provides $\sim$20$\%$ of the peak. 

The measured values of the QPs-boson couplings can be used to estimate the upper bound for $T_c$, expected for each subset of the bosonic function. The maximal critical temperatures  attainable are calculated assuming that $\Pi_b(\Omega)$ entirely contributes to the $d$-wave pairing and fixing the pseudopotential $\mu^*$=0. In the strong-coupling formalism, the critical temperature for $d$-wave pairing in a Fermi liquid with $\mu^*$=0, is approximately given by \cite{Allen1975}:
\begin{equation}
T_{c}=0.83\tilde{\Omega}\exp[-1.04(1+\lambda_i)/g\lambda_i]
\end{equation}
where ln$\tilde{\Omega}$=2/$\lambda_i$$\int_0^{\infty}\Pi_b(\Omega)\mathrm{ln}\Omega /\Omega d\Omega$, $\lambda_i$=$2\int\Pi_{i}\left(\Omega\right)/\Omega\;d\Omega$ is the electron-boson coupling constant and $g$$\in$[0,1] is a parameter that accounts for the $d$-wave nature of the superconducting gap. The upper bound $g$=1 is used in the following.\\
We summarize the main results:\\
-The coupling to bosons of electronic origin is $\lambda_{e-be}$=$2\int I^2 \chi(\Omega)/\Omega\;d\Omega$=\textbf{1.1$\pm$0.2}. This value corresponds to a maximal $T_c$=105-135 K.\\
-The coupling to strongly-coupled phonons (most likely breathing and buckling Cu-O optical modes) is $\lambda_{e-SCP}$=$2\int\alpha^2F(\Omega)_{SCP}/\Omega\;d\Omega$=\textbf{0.4$\pm$0.2}. This value is in complete agreement with the values measured on similar materials via different techniques, such as time-resolved photoemission spectroscopy \cite{Perfetti2007}, time-resolved electron diffraction \cite{Carbone2008} and single-color high-resolution time-resolved reflectivity \cite{Gadermaier2010}. The maximal critical temperature estimated considering only SCPs as the glue is $T_c$=2-30 K.\\
-The coupling to all other phonons is $\lambda_{e-lat}$=$2\int\alpha^2F(\Omega)_{lat}/\Omega\;d\Omega$=\textbf{0.2$\pm$0.2}, corresponding to $T_c$=0-12 K.\\
The large error bars in the values of $\lambda_{e-be}$, $\lambda_{e-SCP}$ and $\lambda_{e-lat}$ include: i) the experimental uncertainty in the pump fluence and other experimental parameters; ii) the possibility of adding to the 4TM a term that accounts for the anharmonic coupling of $SCP$ to the lattice; iii) the possible overestimation of $\lambda_{SCP}$ and $\lambda_{lat}$ by a factor 8/5 \cite{Kabanov2008}, in the extreme case that the electron-electron scattering time ($\tau_{e-e}$) is much larger than the electron-phonon scattering time and the energy exchange between the QPs and bosons begins before the establishment of a quasi-equilibrium fermionic population at $T_e+\delta T_e$.\\

Non-equilibrium optical spectroscopy allows disentangling the electronic and phononic contributions to the total bosonic function $\Pi(\Omega)$ of a cuprate superconductor. The strength of the coupling to bosons of electronic origin and its spectral distribution can account alone for the high-critical temperature ($T_c$=96 K) of the system. This supports the description of high-temperature superconductivity in cuprates in terms of a generalized Eliashberg formalism, in which the attractive interaction is mediated by virtual bosonic excitations of electronic origin, such as spin fluctuations or current loops.

\section{Opening of a pseudogap and failure of the extended Drude model}
In the calculation of the self-energy in Section 2.3, a constant density of states at the Fermi level has been considered. Although this approximation is reliable at T=300 K in optimally and overdoped systems, it fails dramatically  as the temperature and the doping decrease and a pseudogap opens in the electronic density of states.
A further evolution of the EDM, accounting for a non-constant electronic density of states, has been developed by Sharapov and Carbotte \cite{Sharapov2005}, and has been used to analyze spectroscopic data at equilibrium \cite{Hwang2011}.
In this model, the imaginary part of the electronic self energy is given by:
\begin{equation}\label{ESelf_EDM3}
\begin{split}
\mathrm{Im}\Sigma(\omega,T)&=-\pi \int_0^\infty \Pi(\Omega)\{\tilde{N}(\omega+\Omega,T) \left[n(\Omega,T)+f(\omega+\Omega,T)\right]+ \\
&+\tilde{N}(\omega-\Omega,T) \left[1+n(\Omega,T)-f(\omega-\Omega,T)\right]\}         d\Omega
\end{split}
\end{equation}
while Re$\Sigma(\omega,T)$ can be calculated through the Kramers-Kronig relations. To mimic the opening of a gap, the normalized density of states $\tilde{N}(\omega,T)$ can be written in the following form \cite{Hwang2011}:
\\
\begin{equation}\label{N_EDM3}
	\tilde{N}(\omega,T)=\begin{cases} \tilde{N}(0,T)+ [ 1-\tilde{N}(0,T) ] \left( \frac{\omega}{\Delta_{pg}}\right)^2 & for \left| \omega \right| \leqslant \Delta_{pg} \\
1+\frac{2}{3}[ 1-\tilde{N}(0,T)] & for \left| \omega \right| \in (\Delta_{pg}, 2\Delta_{pg}) \\
1 & for \left| \omega \right| \geqslant 2\Delta_{pg} \end{cases}
\end{equation}
\\
Where $\Delta_{pg}$ is the energy gap width, while $\tilde{N}(0,T)$ is the gap filling. In the top-right inset of Figure 5 we report $\tilde{N}(\omega,T)$ for the values $\Delta_{pg}$=40 meV and $\tilde{N}(0,T)$=0.7.
\begin{figure}[t]
\begin{center}
\includegraphics[bb= 280 10 620 380, width=12pc]{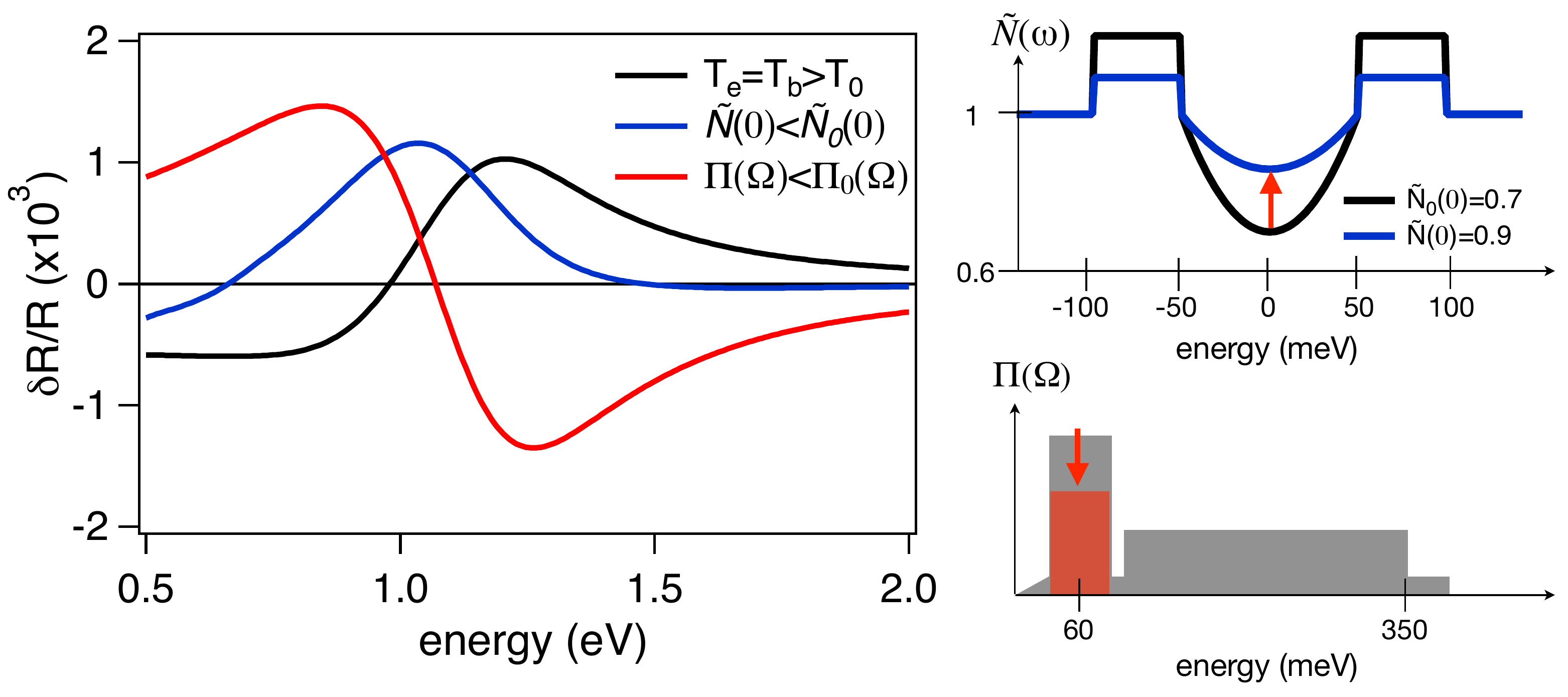}\hspace{2pc}%
\caption{\label{pseudogap}$\delta R/R(\omega,t$=$0)$ calculated in the case of: impulsive heating of the fermonic quasiparticles and bosonic excitations (black line), impulsive filling of the pseudogap (blue line), impulsive quench of the electron-boson coupling (red line). The right insets qualitatively show the change of the density of states, $\tilde{N}(\omega)$, and of the total glue, $\Pi(\Omega)$, assumed in the calculation of $\delta R/R(\omega,t$=$0)$. The equilibrium temperature $T_0$=100 K and the same $\Pi(\Omega)$ as the one determined at T=300 K have been used.}
\end{center}
\end{figure}

Replacing Eqs. 9 and 11 with Eqs. 19 and 20 it is possible to calculate the $\delta R/R(\omega,t)$ induced by the pump pulse, even in the presence of a pseudogap.\\ 
In particular, three different relevant processes can be analyzed:\\
a) the impulsive heating of bosons at a temperature $T_b$ higher than the equilibrium temperature $T_0$ of the system. A quasi-thermal scenario, with $T_e$$\simeq$$T_b$, and $T_0$=100 K is assumed. The calculated $\delta R/R(\omega,t=0)$, reported in Figure 5 (black line), is very similar to the reflectivity variation obtained in the case of a constant density of states at T=300 K (see Figure 4) and is related to the impulsive increase of the scattering rate and broadening of the Drude peak;\\
b) the impulsive filling of the pseudogap, as a consequence of the photoinjection of excess excitations. The $\delta R/R(\omega,t=0)$ is calculated assuming that the value $\tilde{N}$(0,T) is impulsively quenched from 0.7 to 0.9. The corresponding change in the density of states is shown in the top-right inset of Fig. 5. The result (see blue curve in Figure 5) is very different from the $\delta R/R(\omega,t=0)$ expected for an impulsive heating of the bosons. Very similar results are obtained by assuming the closing of the gap ($\Delta_{pg}$$<$$\Delta_{pg0}$) instead of its filling;\\
c) the impulsive decrease of the total $\Pi(\Omega)$. Although $\alpha^2F(\Omega)_{SCP}$ and $\alpha^2F(\Omega)_{lat}$ are expected to be doping- and temperature-independent, $I^2\chi(\Omega)$ could increase as the doping and the temperature decrease because either new magnetic excitations coupled with QPs emerge in the pseudogap phase \cite{Li2010,Li2012} or electronic correlations induce a non-Fermi liquid like increase of the self-energy at the antinodes. $\delta R/R(\omega,t=0)$ is calculated by assuming a small quench of the peak at 60 meV of the bosonic function (bottom-right inset of Fig. 5). The result (red solid line) is reported in Figure 5 and can be rationalized in terms of an impulsive decrease of the coupling and consequent narrowing of the Drude peak, that is, the opposite effect to the transient heating of bosons.

In conclusion, we have shown that the use of non-equilibrium optical spectroscopy can be extended to doping and temperature regimes in which a pseudogap opens and electronic correlations strongly affect the electronic properties of the system.

\section{Conclusions and Perspectives}
Non-equilibrium optical spectroscopy is important and effective for investigating the physics of the cuprate superconductors and, more in general, of strongly correlated electron materials. Adding the temporal degree of freedom to the frequency resolution of equilibrium optical spectroscopy, allows disentangling the electronic and phononic contributions to the Eliashberg \textit{Bosonic Function} $\Pi(\Omega)$ in a prototypical cuprate, i.e. optimally-doped Y-Bi2212. Our results demonstrate that, in principle, the strength of the coupling of quasiparticles to bosons of electronic origin can account for the high-critical temperature of the system ($T_c$=96 K). This result supports a description of high-temperature superconductivity in cuprates in terms of an attractive interaction, mediated by virtual bosonic excitations of electronic origin, where the most prominent candidates are spin fluctuations and loop currents. Furthermore, the analysis of the data obtained from non-equilibrium optical spectroscopy relies on a formalism (the Extended Drude model) that can be easily extended to more complex cases, e.g., systems with a gap in the density of states. This is very promising for further experiments aimed at investigating the elusive nature of the region of the cuprate phase diagram, in which a "pseudogap" in the density of states is present. Finally, the techniques described here can be easily extended to other complex materials, such as different classes of transition-metal oxides and iron-based superconductors.

These new experimental and theoretical tools, developed to investigate the electronic properties of correlated materials out of equilibrium, open intriguing perspectives for the next-years materials science. The modeling of the interaction of ultrashort light pulses with correlated materials is still at its infancy. The use of the "effective" temperatures and of the extended Drude models to describe the non-equilibrium dynamics on the sub-ps timescale has been proved to be a very effective tool to carry out a quantitative analysis of the data and single out the different contributions to the pump-induced modification of the dielectric function. Nonetheless, these approximations cannot account for other important processes that play an important role in the femtosecond dynamics of strongly correlated materials.
In particular: \\
-the origin of the interplay between high-energy Mott-like excitations at 1.5-2 eV and the onset of high-temperature superconductivity, recently demonstrated by non-equilibrium optical spectroscopy \cite{Giannetti2011}, is still an open question directly related to the nature of the superconducting phase in cuprates.\\
-the \textbf{k}-space distribution of the nascent population, after interaction with a femtosecond light pulse, can have strongly non-thermal features, such as an excess of excitations at the antinodes \cite{Cortes2011}. This feature could provide the access to novel non-thermal phases \cite{Kusar2008,Giannetti2009,Coslovich2011,Stojchevska2011,Fausti2011}, in which superconductivity can be controlled by means of light pulses. 

The joint effort of emerging theoretical tools, like non-equilibrium DMFT \cite{Eckstein2009,Amaricci2012} and time-dependent Gutzwiller approach \cite{Schiro2010,Andre2012} and time-resolved experiments probing different physical properties, like time-resolved optical and photoemission spectroscopies and time-resolved electron diffraction, will provide the key to understand the role of the electronic correlations in controlling the ultrafast electronic properties in unconventional superconductors and transition-metal oxides.

\ack
We acknowledge discussions and suggestions from M. Capone and A. Chubukov. C.G., S. D.C. and F.P. have received funding from the European Union, Seventh Framework Programme (FP7 2007-2013), under Grant No. 280555 (GO-FASR project). F.C., G.C., and F.P. acknowledge the support of the Italian Ministry of University and Research under Grant Nos. FIRBRBAP045JF2 and FIRB-RBAP06AWK3. The crystal growth work was performed in M.G.'s prior laboratory at Stanford University, Stanford, CA 94305, USA, and supported by DOE under Contract No. DE-AC03-76SF00515. The work at UBC was supported by the Killam, Sloan Foundation, CRC, and NSERC's Steacie Fellowship Programs (A. D.), NSERC, CFI, CIFAR Quantum Materials, and BCSI. D. v.d.M. acknowledges the support of the Swiss National Science Foundation under  Grant No. 200020-130052 and MaNEP.

\end{document}